\documentclass[preprint2]{aastex}

\shorttitle{Nightside Scintillation with LOFAR}
\shortauthors{Fallows et al.}

\begin{document}

\title{Separating Nightside Interplanetary and Ionospheric Scintillation with LOFAR}

\author{R.A. Fallows}
\affil{ASTRON - the Netherlands Institute for Radio Astronomy, Postbus 2, 7990 AA Dwingeloo, the Netherlands}
\email{fallows@astron.nl}

\author{M.M. Bisi}
\affil{RAL Space, Science \& Technology Facilities Council - Rutherford Appleton Laboratory, Harwell, Oxford, Oxfordshire, OX11 0QX, United Kingdom}
\email{Mario.Bisi@stfc.ac.uk}

\author{B. Forte}
\affil{Dept of Electronic and Electrical Engineering, University of Bath, Bath, BA2 7AY, United Kingdom}
\email{b.forte@bath.ac.uk}

\author{Th. Ulich}
\affil{Sodankyl\"a Geophysical Observatory, T\"ahtel\"antie 62, FIN-99600 Sodankyl\"a, Finland}

\author{A.A. Konovalenko}
\affil{Institute of Radio Astronomy, 4 Chervonopraporna str., 61002 Kharkov, Ukraine} 

\author{G. Mann}
\affil{Leibniz-Institut für Astrophysik Potsdam, An der Sternwarte 16, 14482 Potsdam, Germany}

\and

\author{C. Vocks}
\affil{Leibniz-Institut für Astrophysik Potsdam, An der Sternwarte 16, 14482 Potsdam, Germany}

\begin{abstract}
Observation of interplanetary scintillation (IPS) beyond Earth-orbit can be challenging due to the necessity to use low radio frequencies at which scintillation due to the ionosphere could confuse the interplanetary contribution.  A recent paper by Kaplan {\it et al} (2015) presenting observations using the Murchison Widefield Array (MWA) reports evidence of night-side IPS on two radio sources within their field of view.  However, the low time cadence of 2\,s used might be expected to average out the IPS signal, resulting in the reasonable assumption that the scintillation is more likely to be ionospheric in origin.  To verify or otherwise this assumption, this letter uses observations of IPS taken at a high time cadence using the Low Frequency Array (LOFAR).  Averaging these to the same as the MWA observations, we demonstrate that the MWA result is consistent with IPS, although some contribution from the ionosphere cannot be ruled out.  These LOFAR observations represent the first of night-side IPS using LOFAR, with solar wind speeds consistent with a slow solar wind stream in one observation and a CME expecting to be observed in another.
\end{abstract}


\section{Introduction}
\label{sec:intro}

The use of interplanetary scintillation (IPS - \citet{Clarke:1964}, published by \citet{Hewishetal:1964}) to observe the solar wind beyond Earth-orbit can be a challenging proposition with few papers dedicated to the subject.  Early papers described observations of the level of scintillation of B0531+21 out to 180$^{\circ}$ from the Sun \citep[e.g.][]{ArmstrongColes:1978}.  More recently, the Ukrainian URAN and UTR-2 telescopes have been used to estimate solar wind speeds beyond Earth orbit from observations of IPS \citep[e.g.][]{Falkovichetal:2010, Olyak:2013}.  One of the challenges is the necessity to use low radio frequencies where the ionosphere could be the dominant source of any scintillation seen.  Regular observations of IPS inside of Earth-orbit, by contrast, are usually taken during local daytime hours and observatories such as the Institute for Space-Earth Environmental Research (ISEE), Japan, \citep[e.g.][]{kojimaKakinuma:1987}, and Ooty, India \citep[e.g.][]{ManoharanAnanth:1990}, IPS arrays use a higher observing frequency. 

In a recent letter, \citet{Kaplanetal:2015}, hereinafter referred to as K2015, presented wide-field ``snapshot'' imaging observations using the Murchison Widefield Array (MWA; \citet{Lonsdaleetal:2009} and \citet{Tingayetal:2013}) in which they claimed to see, from successive images, IPS on flux measurements of two sources within the field of view, despite a 2\,s time cadence between images which might be expected to average out the IPS signal.  The observations were also taken at night, with the scintillating sources at solar elongations of $\sim$110-115$^{\circ}$, potentially indicating that the scintillation seen could be ionospheric in origin.  Hence, the question arises whether or not the interplanetary medium is the dominant source of the stochastic variations seen in the received signal.  K2015 goes into significant detail to allay concerns, but the current lack of a high time-cadence capability (although post-processing of voltage-capture data is now underway) does not allow for a proper evaluation of the scintillation seen.  The use of high time-cadence observations can help to ascertain the combination of IPS and ionospheric scintillation contributions to the observed signal intensities.

The Low Frequency Array (LOFAR; \citet{LOFAR-reference-paper:2013}), a modern radio telescope based in the Netherlands but with a number of stations across Europe is capable of observing frequencies in the range 10--250\,MHz, including full coverage of those used by K2015.  It has on-line beam-forming capabilities and the ability to record data per station, enabling it to be used as a large collection of individual telescopes, with baselines ranging from $\sim$50 metres to $\sim$1,500 kilometres (as of early 2016), in similar fashion to more-traditional systems.  Several observations of IPS have been carried out using LOFAR since full operations commenced in 2012 (initial observations are presented in \citet{Fallowsetal:2012a, Bisietal:2016}) and irregular monitoring of ionospheric scintillation has been performed since 2014 (Fallows {\it et al.} in prep.). 

The Kilpisj\"arvi Atmospheric Imaging Receiver Array (KAIRA; \citet{KAIRA-reference-paper:2014}), a station built using LOFAR hardware in arctic Finland, has been routinely monitoring the ionosphere, including ionospheric scintillation, since 2012 \citep[e.g.][]{Fallowsetal:2014}: The ionospheric scintillation conditions above KAIRA are naturally more severe than above LOFAR: At auroral latitudes, refractive index gradients due to field-line elongated ionisation structures are stronger than in the case of middle latitudes structures.  These observations can, therefore, be used to verify the effects of periods of strong ionospheric scintillation. 

In this letter, we use observations of interplanetary and ionospheric scintillation from both of these arrays to provide a comparison with the K2015 result.

\section{Observations and Results}
\label{sec:observations}

The observations presented here are analysed with the aim of answering three specific questions:-

\begin{itemize}
	\item Is IPS averaged out with an integration time of 2\,s?
	\item Is IPS observed beyond Earth-orbit, and could it be confused with ionospheric scintillation?
	\item Which power spectra, those from IPS or those from ionospheric scintillation, are more consistent with the K2015 result?
\end{itemize}

In November 2015 a series of observations were taken under an ionospheric scintillation monitoring project, LC5\_001, to observe both 3C48, a very compact source known as one of the strongest scintillators from plasma structures in the interplanetary medium, and Cassiopeia A, a relatively broad source known to scintillate at low radio frequencies from plasma structures in the ionosphere, but too broad to scintillate from plasma structures in the interplanetary medium.  LOFAR was set up to record beam-formed data from each station individually (``Fly's Eye'' mode - see \citet{Stappersetal:2011}) over the frequency range 110--178\,MHz, with a frequency resolution of 12\,kHz and a time cadence of approximately 0.01\,s.  The data were averaged in post-processing to a final frequency resolution of 195\,kHz and time resolution of approximately 0.1\,s.  The stations of the LOFAR ``core'', a dense group of stations covering an area with a diameter of approximately 3\,km, were used to observe Cassiopeia A; remaining stations across the Netherlands and internationally were used to observe 3C48.

\begin{figure}
	\centering
	\includegraphics[width=7cm]{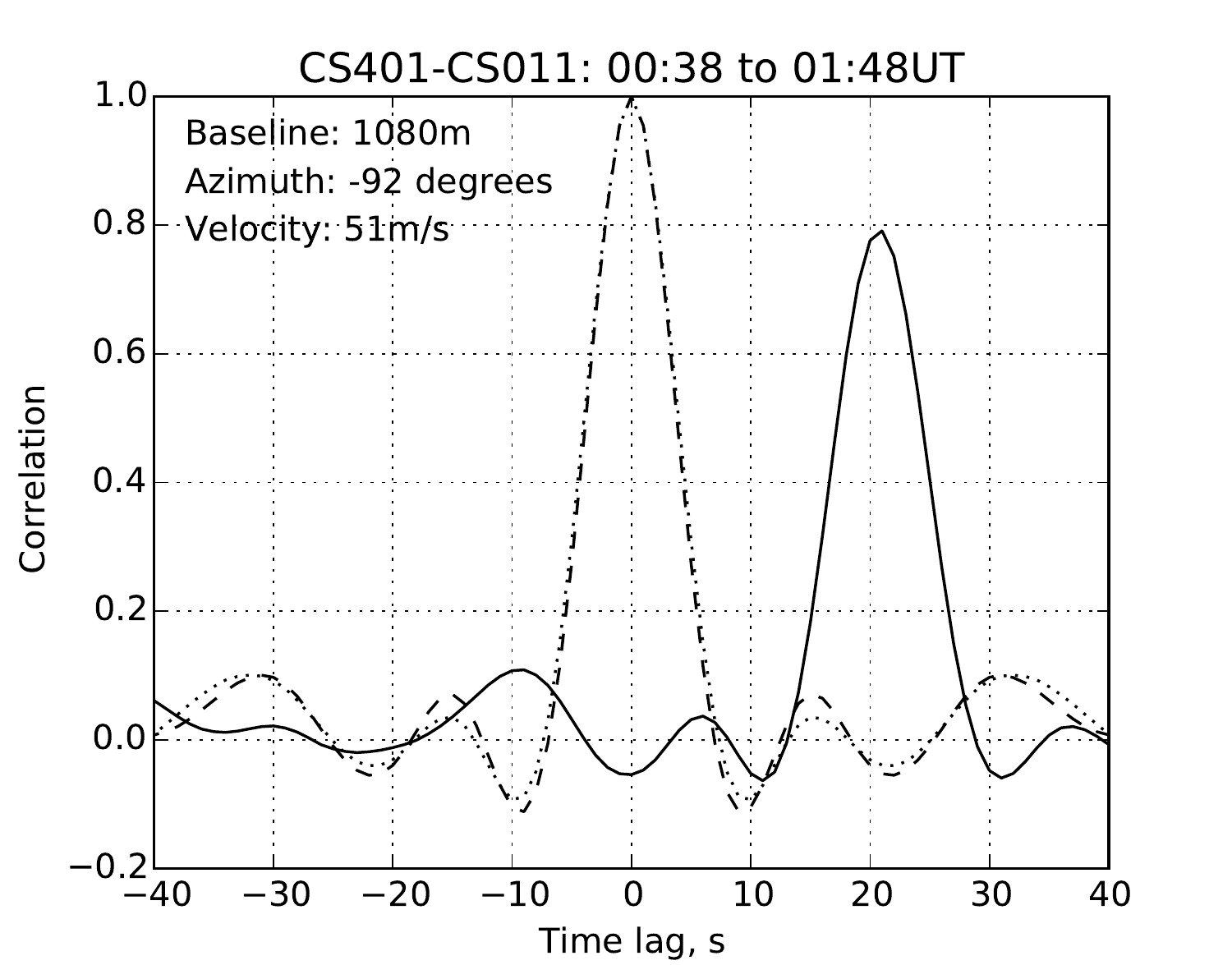}
	\includegraphics[width=7cm]{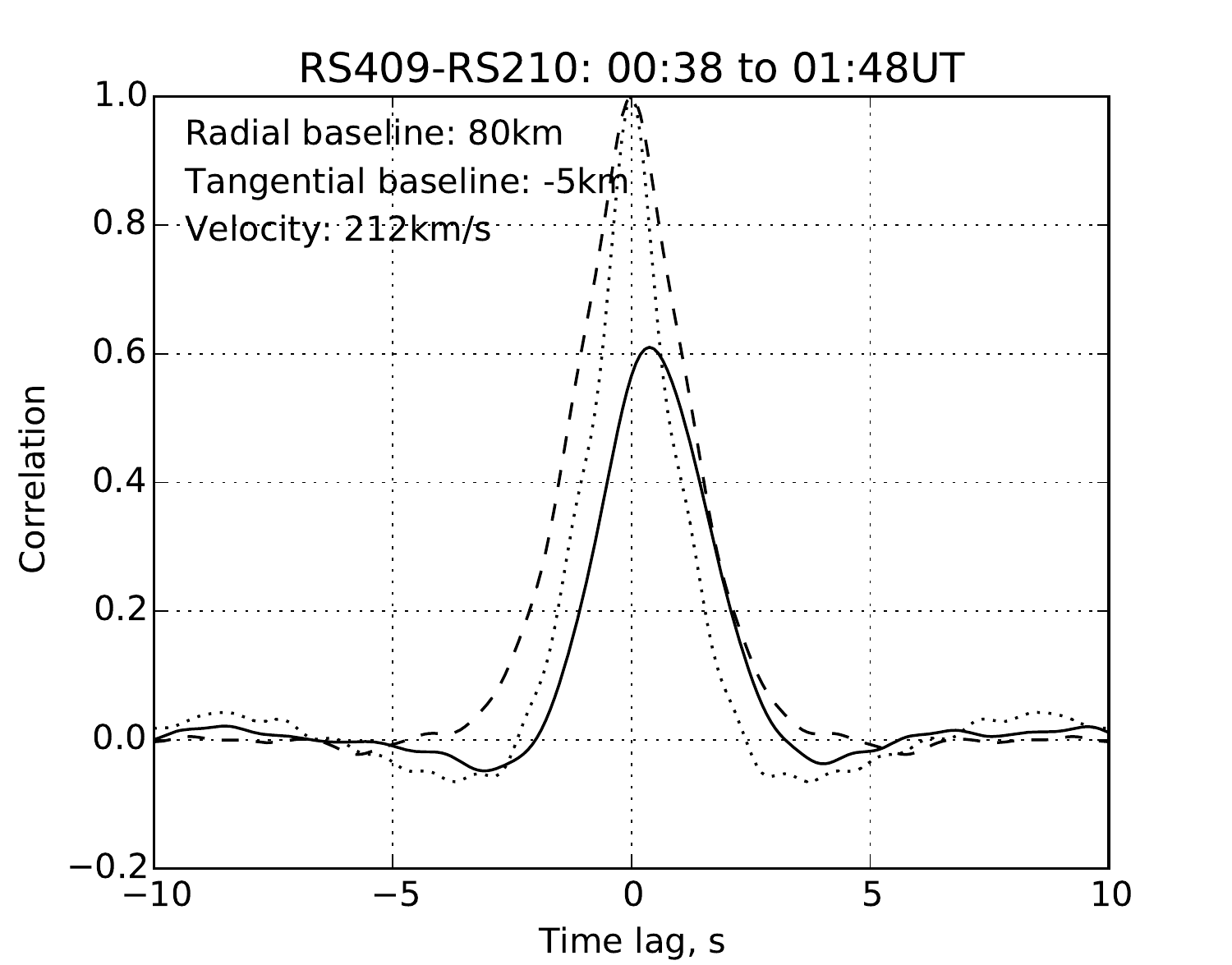}
	\includegraphics[width=7cm]{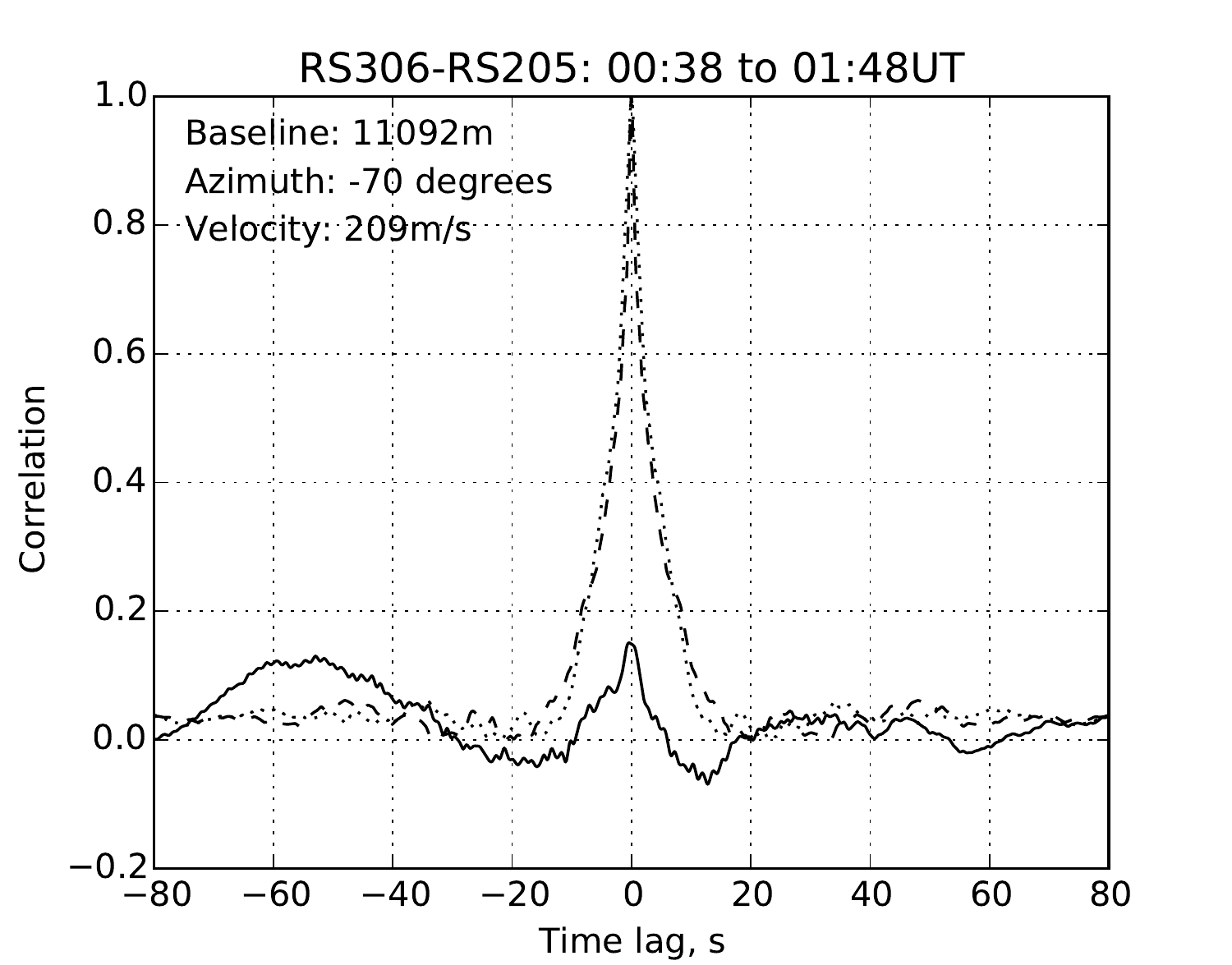}
	\caption{Plots of auto- (dashed and dotted lines) and cross-correlation (solid line) functions of time series' calculated from the observations of 8 November 2015, over the entire duration of the observations.  Top: Cassiopeia A data from core stations CS401 and CS011; middle: 3C48 data from remote stations RS409 and RS210; bottom: 3C48 data from remote stations RS306 and RS205.}
	\label{fig:comparetimeseries}
\end{figure}

At this time, 3C48 was at a solar elongation of approximately 157$^{\circ}$ and scintillation was evident upon inspecting the data.  This is a greater elongation than the K2015 observations and any IPS is expected to be weaker as a consequence.  The origin of the 3C48 scintillation is confirmed using a cross-correlation analysis:  In the case of the ionosphere, bulk flows of 10s to 100s of metres per second lead to a time delay of several, and possibly 10s, of seconds over the short baselines between stations within the LOFAR core (for baselines with a component aligned with the ionospheric bulk flow).  The solar wind flows much faster and even a slow solar wind stream of approximately 350\,km\,s$^{-1}$ leads to time delays of less than a second between any pair of LOFAR remote stations, with baseline lengths of tens of kilometers.  Correlation of IPS is also expected over international station baselines of hundreds of kilometers.  

In order to calculate power spectra and correlation functions, time series' were first obtained by taking the median over the pass-band of interest from the data received by each station.  To match the data presented in K2015, only 32\,MHz of the recorded bandwidth was used, centred on 155\,MHz.  A threshold was also applied to the time series' to clip obvious spikes due to radio frequency interference (RFI).  Power spectra were calculated using Welch's method, averaging spectra with a 50\% overlap; cross-spectra were calculated between station-pairs using the same method.  For calculation of the correlation functions, high- and low-pass filters were applied to the spectra to remove slow system variations and white noise respectively.

We present two sets of observations: one taken on 8 November 2015 and the other on 10 November 2015.

\subsection{3C48 and Cassiopeia A on 8 November 2015}
\label{subsec:8November2015}

This observation ran from 00:38 to 01:48\,UT, with observation IDs L403712 and L403714 for 3C48 and Cassiopeia A respectively.  Due to an erroneous setup for the observation of Cassiopeia A, these data have the lower time resolution of 1\,s.  A weak impact from a Coronal Mass Ejection (CME) was recorded by the Advanced Composition Explorer (ACE) spacecraft in the early evening of 6 November 2015.  The speed recorded by ACE was around 560\,km\,s$^{-1}$, rising to $\sim$700\,km\,s$^{-1}$ as the CME progressed.  In the $\sim$30\,hrs between this CME starting to traverse Earth orbit and the time of these observations, it is likely to have travelled a further $\sim$0.4\,AU with the material predominantly off the same side of Earth as the line of sight to 3C48.  Hence it is highly likely that the line of sight passed through a portion of this CME at the time, making a further suitable comparison with the assumptions made by K2015.

\begin{figure}[h]
	\centering
	\includegraphics[width=8cm]{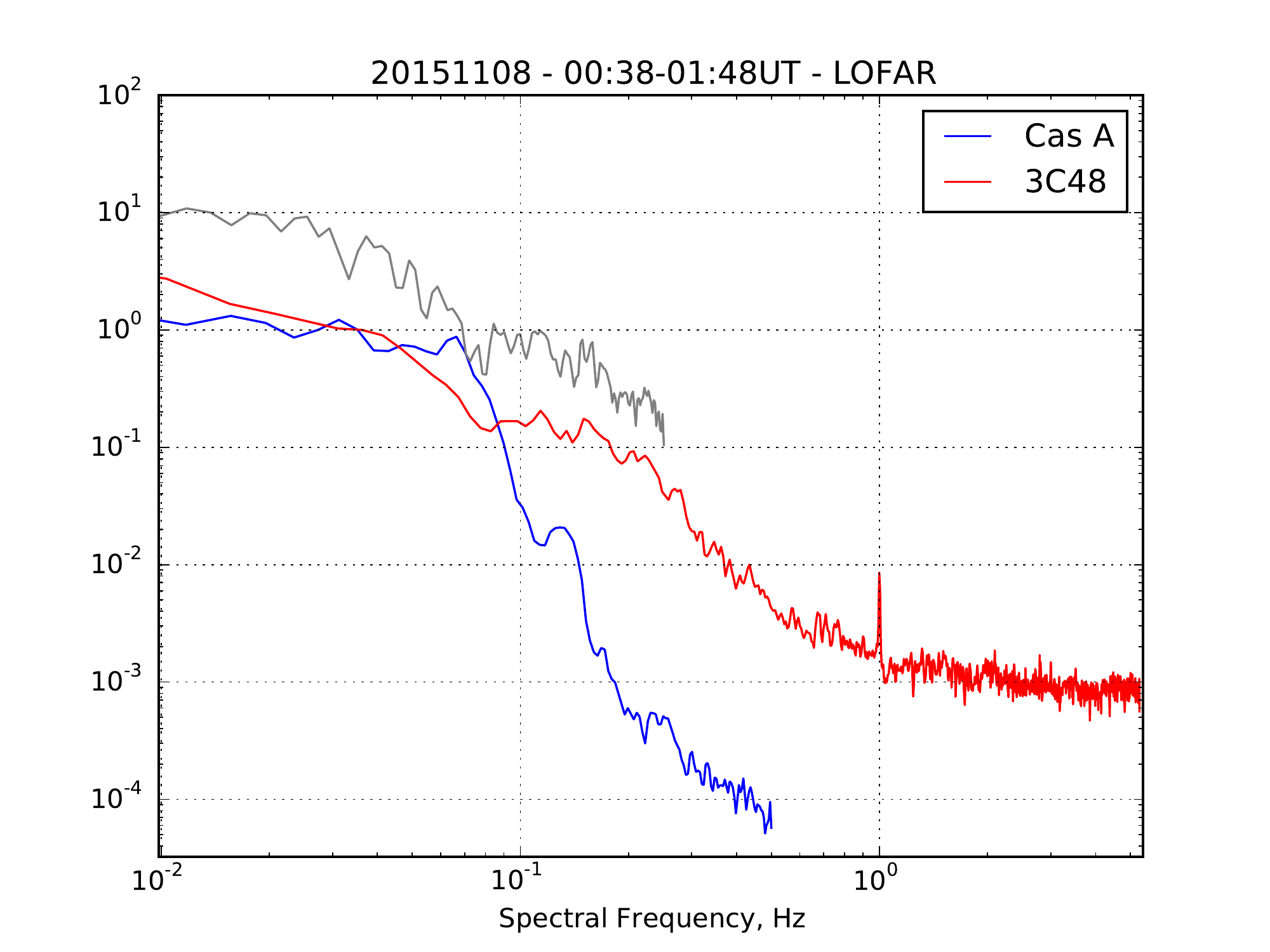}
	\caption{Power spectra: 3C48 data from remote station RS406 are plotted in red; Cassiopeia A data from core station CS026 are plotted in blue. For ease of comparison these spectra were normalised such that the level is matched at the low spectral frequencies.  Also plotted in grey, but shifted upwards so that it does not confuse the other spectra, is a power spectrum of the 3C48 data averaged to a 2\,s cadence.}
	\label{fig:comparespectra}
\end{figure}

Example correlation functions are presented in Figure \ref{fig:comparetimeseries}. The Cassiopeia A data show high correlation with a time delay of $\sim$20\,s on a baseline of 1.08\,km, equating to a drift speed of $\sim$51\,m\,s$^{-1}$.  This should not be taken as a direct measurement of the ionospheric drift speed: the correlations presented are examples only and the baselines used may not be exactly aligned with the drift direction.

The middle plot of Figure \ref{fig:comparetimeseries} shows the cross-correlation function (CCF) of 3C48 data between remote stations RS409 and RS210. The high-pass filter used was applied at a spectral frequency of 0.08\,Hz.  The CCF has a time delay of $\sim$0.38\,s on a baseline of 80\,km which was approximately aligned with the solar wind outflow, indicating a drift speed of $\sim$212\,km\,s$^{-1}$.  This is clearly inconsistent with speeds expected from the ionosphere but much lower than the speed expected from IPS.  The drift observed is perpendicular to the line of sight between radio source and Earth; inside of Earth-orbit, the scintillation pattern observed can be assumed to be mostly the result of scattering around the point of closest approach of the line of sight to the Sun, and the solar wind, assumed to be radial in direction, flows perpendicular to it at this point.  For observations beyond Earth-orbit, the solar wind flow is nowhere perpendicular to the line of sight and the IPS drift speed represents a foreshortening of the solar wind speed.  The angle between the solar radial direction and the measured IPS velocity for a line of sight with an elongation of 157$^{\circ}$ is 67$^{\circ}$ if measured at Earth, leading to a corrected velocity of 542\,km\,s$^{-1}$.  This calculation assumes minimum foreshortening and so a minimum velocity, but is consistent with the speed expected from the CME measurement by ACE. 

The lower plot of Figure \ref{fig:comparetimeseries} shows the CCF of 3C48 data between RS306 and RS205; in this case the high-pass filter was applied at the lower spectral frequency of 0.02\,Hz to better show slower time variations which may correspond to any ionospheric component.  The CCF indicates a lower correlation near zero time lag (which corresponds with the IPS correlation seen in the middle plot), but also a low, but significant, correlation at a time lag of $\sim$-56\,s.  The baseline between these two stations is relatively short (11\,km) and not well-aligned with the solar wind outflow, which would reduce the correlation due to IPS.  The long-time lag correlation corresponds to a drift speed of $\sim$209\,m\,s$^{-1}$, a speed consistent with those expected in the ionosphere.  Other CCFs from baselines with a similar alignment show similar results, whereas different alignments do not, giving confidence that this correlation is due to an ionospheric component. 

Power spectra for both sources are shown in Figure \ref{fig:comparespectra}, using Welch's method with 2048 points per averaged spectrum for the 3C48 data and 256 points per spectrum for the Cassiopeia A data.  A further spectrum using 2\,s averaged data was calculated from the 3C48 measurement, using 256 points as in K2015.

A sharply-defined Fresnel knee is seen at 0.07\,Hz in the Cassiopeia A spectrum. In the 3C48 spectrum, a knee is evident at around 0.15\,Hz, corresponding to the IPS component.  The Cassiopeia A spectrum shows a steeper decline than that of 3C48, indicating a faster cascade from larger to smaller scales. Comparing the 2\,s 3C48 spectrum with the spectra seen in Figure 3 of K2015, particularly their spectrum of PKS B2318-195, it can be seen that the spectra are broadly similar: both show a slight flattening at the highest spectral frequencies and a slight excess power at the lowest spectral frequencies, inside of ~0.03\,Hz.  The Cassiopeia A spectrum is clearly inconsistent with the spectra of K2015.  This also indicates that the 2\,s time resolution has not completely filtered out the IPS component.

\subsection{3C48 and Cassiopeia A on 10 November 2015}
\label{subsec:13November2015}

This observation ran from 17:05 on 10 November 2015 to 02:45\,UT on 11 November 2015, with observation IDs L403976 and L403980 for 3C48 and Cassiopeia A respectively.  The international stations were available for this observation: these contain twice the number of antennas of the Dutch remote stations with a corresponding increase in sensitivity, and enable longer baselines to be used.  Data were analysed in 30-minute intervals.

\begin{figure}
	\centering
	\includegraphics[width=8cm]{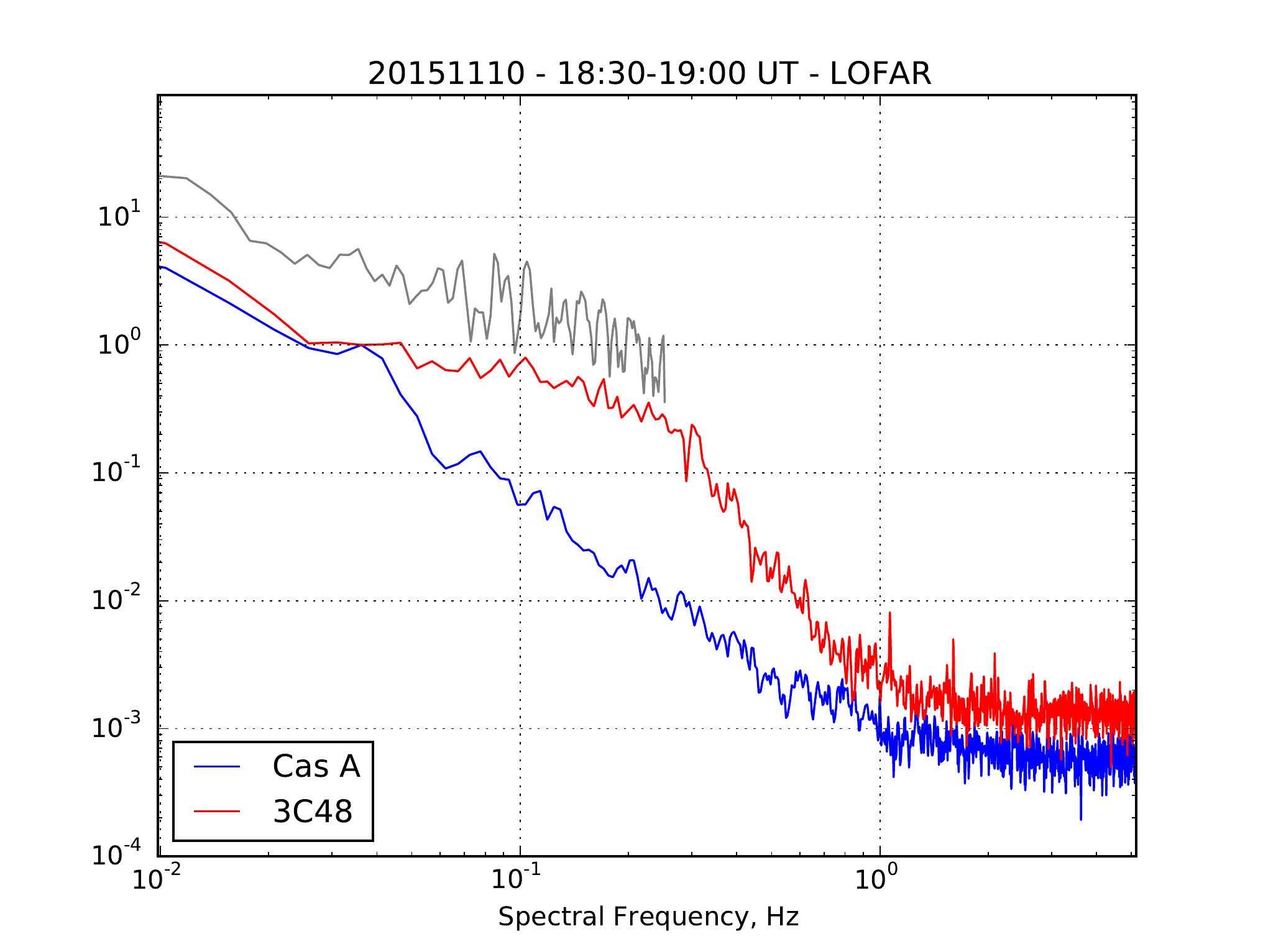}
	\includegraphics[width=8cm]{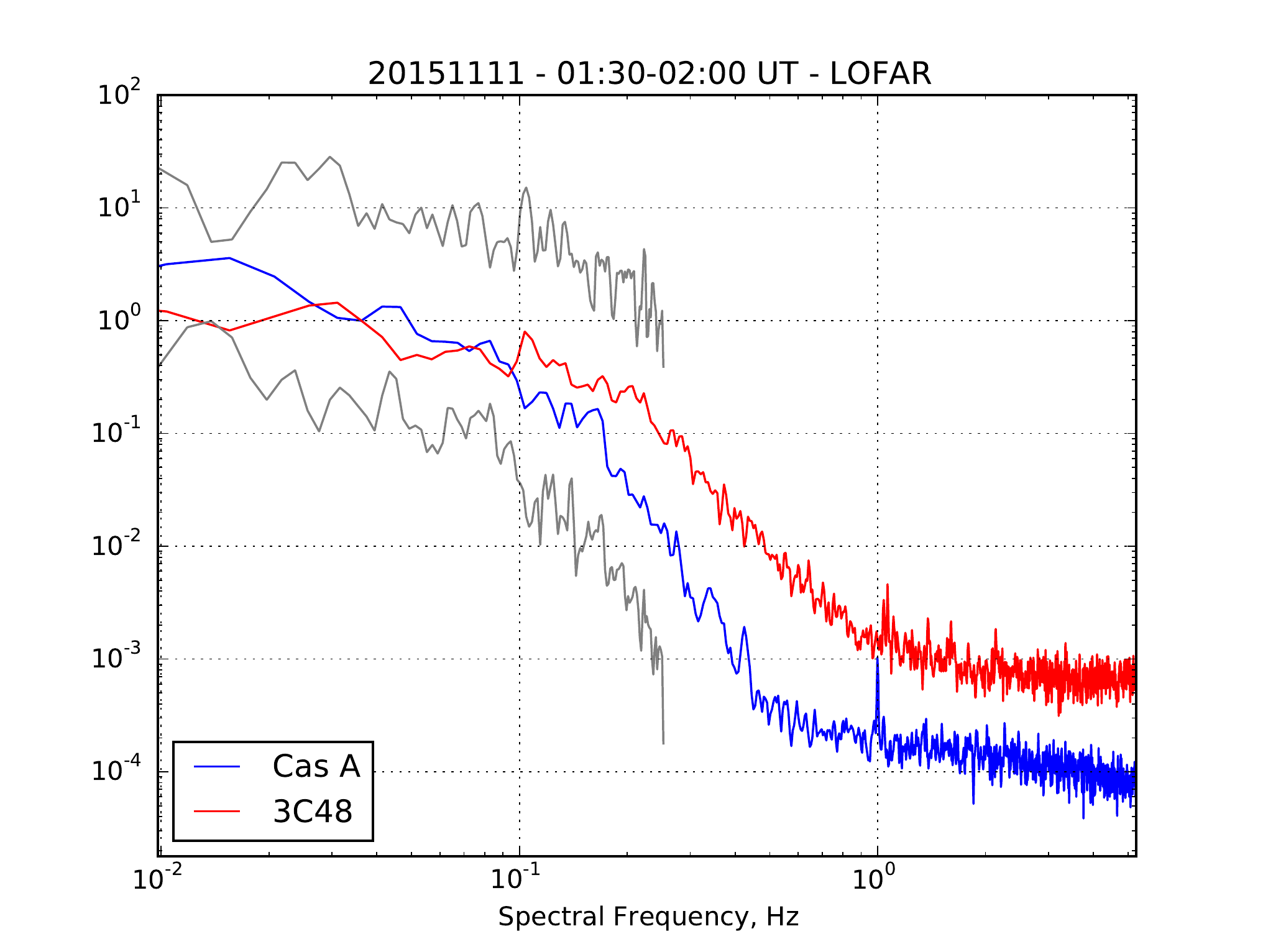}
	\caption{Example power spectra from the observation of 10-11 November 2015: 3C48 data from international station UK608 are plotted in red; Cassiopeia A data from core station CS501 are plotted in blue. For ease of comparison these spectra were normalised such that the level is matched at the low spectral frequencies. Top: spectra from 18:30 to 19:00\,UT; plotted in grey is a power spectrum of the 3C48 data averaged to a 2\,s cadence. Bottom: spectra from 01:30 to 02:00\,UT; plotted in grey are power spectra from both 3C48 (upper) and Cassiopeia A (lower) data averaged to a 2\,s cadence.}
	\label{fig:comparespectra2}
\end{figure}

As with the observation of 8 November, the CCFs confirm that IPS is evident in the 3C48 data.  Figure \ref{fig:comparespectra2} shows power spectra from two 30-minute segments.  The 3C48 spectra show well-defined Fresnel knees around 0.3\,Hz for the earlier time interval and around 0.2\,Hz for the later interval. Spectra from 2\,s averaged data appear consistent with those of K2015.  The Fresnel frequency for the CasA spectrum at 18:30\,UT is lower and distinct from that of 3C48. This is consistent with the likely presence of elongated ionospheric structures originated by particle precipitation in the auroral ionosphere, with typically low ionospheric drift. Later, in the 01:30\,UT spectra, the CasA spectrum is broadened in response to the transition to a stronger scattering regime, with a Fresnel frequency closer to that of 3C48. This is consistent with both the presence of stronger ionisation gradients as well as with typical ExB drift in the nighttime auroral ionosphere.  A spectrum calculated from 2\,s averaged Cassiopeia A data is also presented: the decline in power at the high spectral frequencies of this spectrum appears more consistent with the spectrum of B2322-275 from the comparison night used in K2015 than those thought to be of IPS. 

\begin{figure}[h]
	\centering
	\includegraphics[width=8cm]{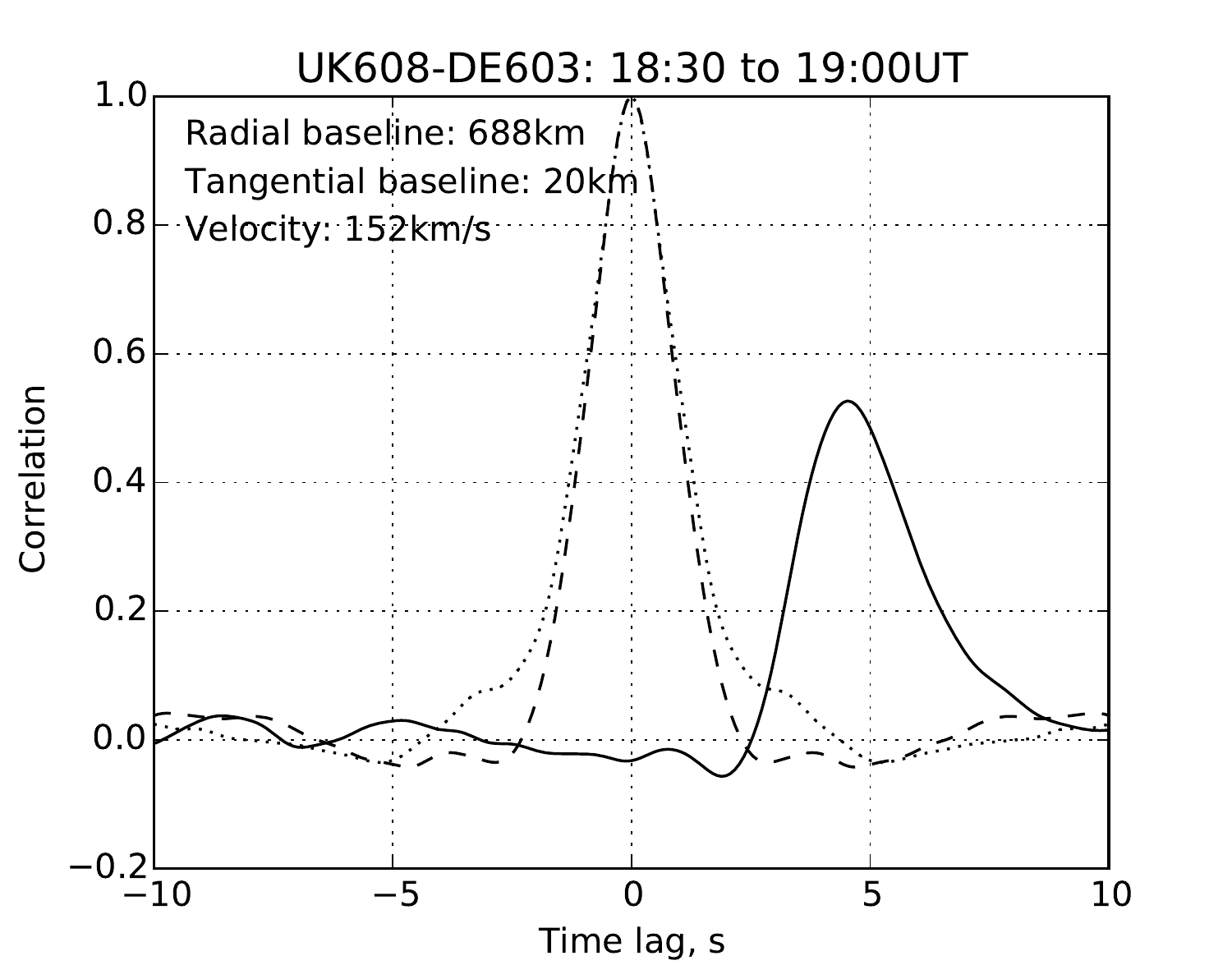}
	\caption{Plots of auto- (dotted and dashed lines with peak valuses of 1.0) and cross-correlation function of time series' calculated from UK608 and DE603 data between 18:30 and 19:00\,UT on 10 November 2015.}
	\label{fig:intcorrelation}
\end{figure}

Figure \ref{fig:intcorrelation} shows the correlation functions of data from UK608 (Chilbolton, UK) and DE603 (Tauntenburg, Germany) from the 18:30\,UT time interval: a clear CCF is seen, giving an estimated solar wind speed of $\sim$152\,km\,s$^{-1}$.  Correcting for foreshortening as before leads to a minimum solar wind velocity of 389\,km\,s$^{-1}$, which corresponds to speeds broadly expected from the slow solar wind.

\section{Comparison with KAIRA}
\label{sec:KAIRA}

\begin{figure}[h!]
    \centering
    \includegraphics[width=8cm]{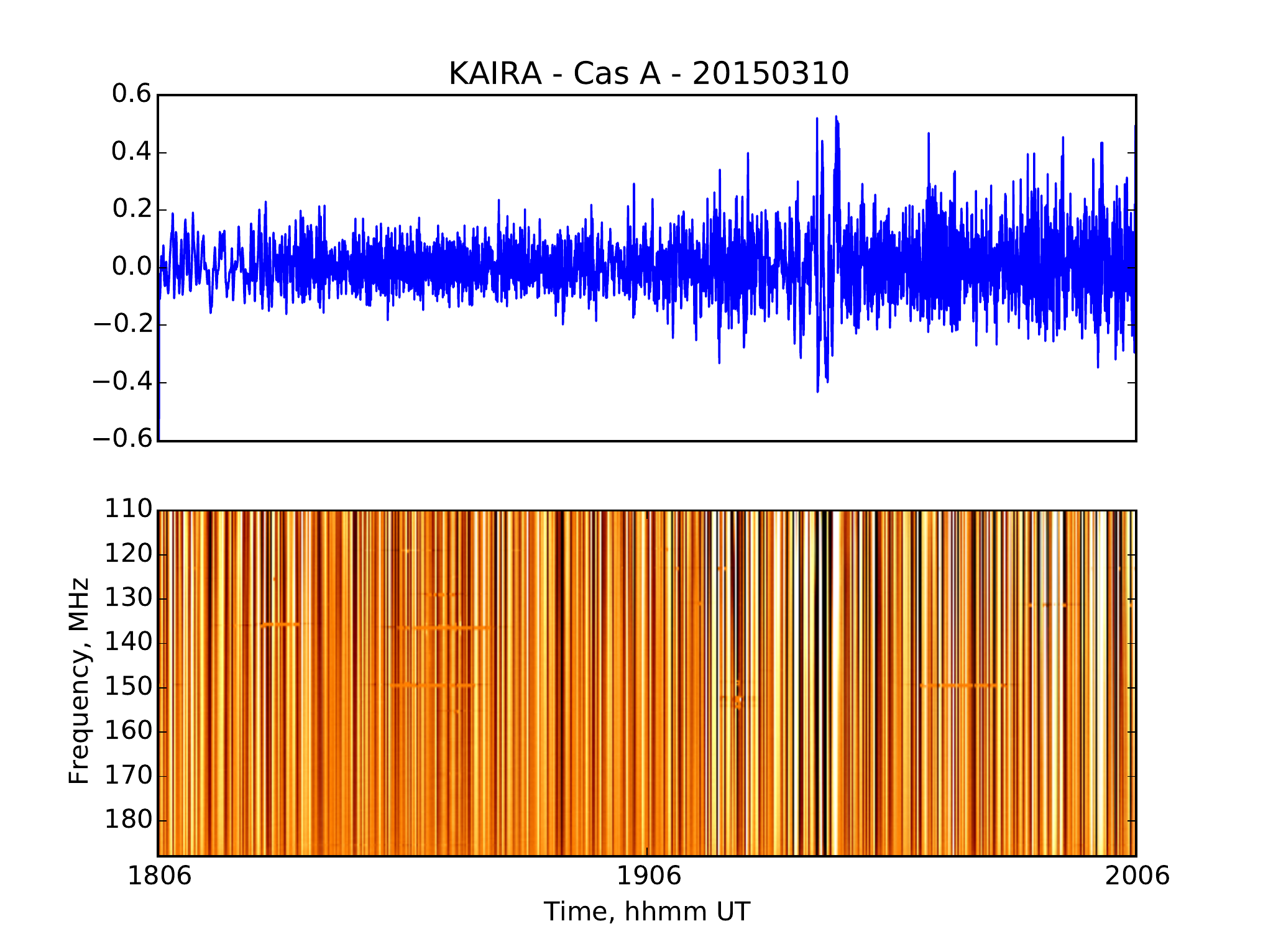}
	\includegraphics[width=8cm]{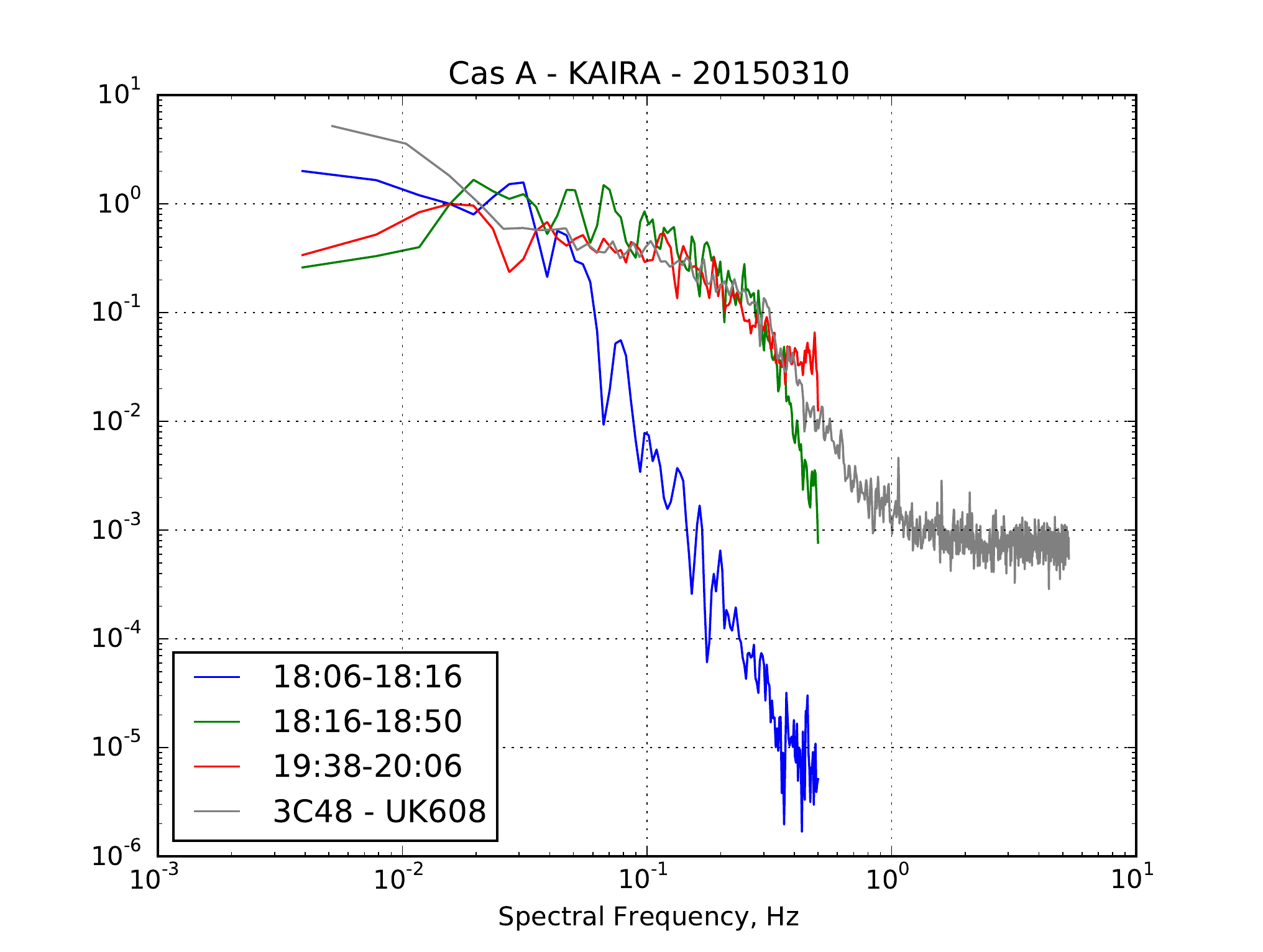}
    \caption{Median time series and dynamic spectrum (upper two plots) across the passband of data from an observation of Cassiopeia A taken using KAIRA on 10 March 2015.  The dynamic spectrum is displayed using the scale -0.02 (dark) to 0.02 (light), arbitrary units.  The lower plot presents example power spectra from different parts of the observation, as shown in the key.  Also shown in grey is the UK608 power spectrum from the 18:30 to 19:00\,UT segment of the LOFAR 3C48 observation of 10 November 2015 for comparison. }
    \label{fig:kairadynspec}
\end{figure}

Further observations of ionospheric scintillation have been taken by the KAIRA station situated in northern Finland.  Its high geomagnetic latitude location means that it is situated under a much more active ionosphere than LOFAR.  An observation taken on 10 March 2015 illustrates the range of conditions seen, as demonstrated in Figure \ref{fig:kairadynspec}.  These data were taken at the lower time resolution of 1\,s.

The time scale of the scintillation varies considerably through the course of this two-hour observation:  The effect of this variation on the power spectrum is also illustrated in Figure \ref{fig:kairadynspec}, where spectra for three sample periods through this observation have been computed.  The power spectrum of UK608 data from 18:30 to 19:00\,UT in the LOFAR 3C48 observation from 10 November 2015 is also shown for comparison.

The power spectrum of the first ten minutes of the KAIRA observation is clearly distinct from the IPS seen in the 3C48 spectrum presented here.  However, the remaining two example power spectra match the 3C48 spectrum almost exactly, illustrating that scintillation from both regimes would be impossible to distinguish from power spectra alone in this instance.

\section{Conclusions}
\label{sec:conclusions} 

The results presented here lead to a few main conclusions:

\begin{itemize}
	\item IPS is not completely averaged out with a 2\,s time resolution;
	\item IPS is observed substantially beyond Earth-orbit with LOFAR, with estimated solar wind speeds consistent with the probable CME in the 8 November 2015 observation, and a slow solar wind stream in the 10 November observation;
	\item The low-cadence IPS power spectra presented here are consistent with those presented in K2015, but also demonstrate that an ionospheric contribution is bound to be present;
	\item The low-cadence LOFAR ionospheric scintillation power spectra presented here, taken under quiet conditions, are not consistent with the scintillation spectra given in K2015;
	\item The KAIRA ionospheric scintillation spectra demonstrate a circumstance under which the two would be indistinguishable from power spectra alone.
\end{itemize}

LOFAR has the advantage of being an array of individual stations which can be used to establish whether observed scintillation is predominantly interplanetary, ionospheric, or a mixture of both.  Under the present setup, MWA does not enjoy this advantage and so establishing which scintillation regime is being observed is dependent on the number of sources scintillating in their field of view and how compact they are:  K2015 state that they only observed scintillation from two more-compact sources in their entire field of view and that ionospheric scintillation would be observed in the majority of sources if it were more prevalent during the observation.  This statement is borne out from LOFAR imaging observations, where any significant ionospheric scintillation is observed throughout the field of view and not on only two sources within it (de Bruyn and others, private communications).  The LOFAR field of view is narrower than that of MWA, but it would still be expected that, at the least, several other sources in the immediate vicinity of the ones exhibiting scintillation in the MWA observation would do so if dominated by the ionosphere, and not only the most-compact two.  This lends further confidence to the conclusions of K2015 that the scintillation they observed was indeed predominantly IPS.  

The time resolution of 2\,s used by K2015 would not allow any reasonable modelling of individal power spectra to obtain solar wind speed or other parameters.  This is illustrated by the LOFAR power spectra from 18:30\,UT on 10 November 2015 given in Figure \ref{fig:comparespectra2}: here the Fresnel knee which we assume to be due to IPS is at a spectral frequency of 0.3\,Hz, beyond the 0.25\,Hz limit of 2\,s time resolution spectra.  However, the K2015 results do raise the question of what could be possible given the ability to do high-time-resolution imaging.

Here, we have also demonstrated that realistic observations of night-side IPS are possible with LOFAR using cross-correlation techniques.  Modelling these results is more challenging as the common assumption of scintillation from around the point of closest approach of the line of sight to the Sun dominating the measurement is invalid once that point becomes the Earth itself.  It may be possible, however, to apply the techniques described by \citet{Falkovichetal:2010} and \citet{Olyak:2013} to LOFAR observations of IPS.  

Finally, this brief investigation has raised further questions about the conditions under which IPS and ionospheric scintillation can be confused. A more-comprehensive study is now underway to look into these, both theoretically and observationally.

\acknowledgments

LOFAR, the Low Frequency Array designed and constructed by ASTRON, has facilities in several countries, that are owned by various parties (each with their own funding sources), and that are collectively operated by the International LOFAR Telescope (ILT) foundation under a joint scientific policy.  KAIRA was funded by the University of Oulu and the FP7 European Regional Development Fund and is operated by Sodankyl\"a Geophysical Observatory. MMB acknowledges Science and Technology Facilities Council (STFC) Core Space Weather funding and also his contribution to this material is based upon work supported by the Air Force Office of Scientific Research, Air Force Material Command, USAF under award number FA9550-16-1-0084DEF. All data are available upon request to the corresponding author. 

Facilities: \facility{LOFAR, KAIRA}.


\end{document}